\documentclass{wqcd03}                 

\usepackage{times}

\usepackage{amssymb}
\usepackage{graphicx}
\def\be{\begin{equation}}
\def\ee{\end{equation}}
\def\bea{\begin{eqnarray}}
\def\eea{\end{eqnarray}}
\def\nnb{\nonumber}
\def\Rate{{\cal R}}

\def\vp{\vphantom{\frac12}}
\def\log{\ln}

\confname{QCD@Work 2003 - International Workshop on QCD, Conversano, Italy, 
14--18 June 2003}

\title{Resummed jet rates with heavy quarks in e$^+$e$^-$ collisions
\thanks{Work supported by the EC 5th Framework Programme under
        contract numbers HPMF-CT-2000-00989 and HPMF-CT-2002-01663.}}

\author{
Germ\'an Rodrigo~\thanks{e-mail: {\tt German.Rodrigo@cern.ch}. Speaker at the Workshop.} and 
Frank Krauss~\thanks{e-mail: {\tt Frank.Krauss@cern.ch}. Present address: Institut 
f\"ur Theoretische Physik, TU Dresden, D-01062 Dresden, Germany.}}

\address{Theory Division, CERN, CH-1211 Geneva 23, Switzerland}

\begin{document}

\begin{abstract}
Expressions for Sudakov form factors for heavy quarks are presented. They 
are used to construct resummed jet rates in $e^+e^-$ annihilation. 
Predictions are given for production of bottom quarks at LEP 
and top quarks at the Linear Collider. 
\end{abstract}

\maketitle

\confname{QCD@Work 2003 - International Workshop on QCD, Conversano, Italy, 
14--18 June 2003}




\section{Introduction}

The formation of jets is the most prominent feature of perturbative QCD in 
$e^+e^-$ annihilation into hadrons. Jets can be visualized as large
portions of hadronic energy or, equivalently, as a set of hadrons confined 
to an angular region in the detector. In the past, this qualitative 
definition was replaced by quantitatively precise schemes to define and 
measure jets, such as the cone algorithms of the Weinberg--Sterman 
\cite{Sterman:1977wj} type or clustering algorithms, e.g. the Jade 
\cite{Bartel:1986ua,Bethke:1988zc} or the Durham scheme ($k_\perp$ scheme) 
\cite{Catani:1991hj}. A refinement of the latter one is provided by the 
Cambridge algorithm \cite{Dokshitzer:1997in}.  
Equipped with a precise jet definition the determination of jet production 
cross sections and their intrinsic properties is one of the traditional 
tools to investigate the structure of the strong interaction and to deduce 
its fundamental parameters. In the past decade, precision measurements, 
especially in $e^+e^-$  annihilation, have established both the gauge 
group structure underlying QCD 
and the running of its coupling constant $\alpha_s$ over a wide range of 
scales. In a similar way, also the quark masses should vary with the scale. 

A typical strategy to determine the mass of, say, the bottom-quark at the 
centre-of-mass (c.m.) energy of the collider is to compare the ratio of 
three-jet production cross sections for heavy and light quarks 
\cite{Rodrigo:1997gy,Abreu:1997ey,Brandenburg:1999nb,Barate:2000ab,Abbiendi:2001tw}. 
At jet resolution scales below the mass of the quark, i.e. for gluons 
emitted by the quark with a relative transverse momentum $k_\perp$ smaller 
than the mass, the collinear divergences are regularized by the quark mass. 
In this region mass effects are enhanced by large logarithms $\log(m_b/k_\perp)$, 
increasing the significance of the measurement. Indeed, this leads to a 
multiscale problem since in this kinematical region also large logarithms 
$\log(\sqrt{s}/k_\perp)$ appear such that both logarithms need to be resummed 
simultaneously.  
A solution to a somewhat similar two-scale problem, namely for 
the average sub-jet multiplicities in two- and three-jet events in $e^+e^-$ 
annihilation was given in~\cite{Catani:1992tm}.
We report here on the resummation of such logarithms
in the $k_\perp$-like jet algorithms~\cite{Krauss:2003cr} and provide some 
predictions for heavy quark production. A preliminary comparison with 
next-to-leading order calculations of the three-jet 
rate~\cite{Rodrigo:1997gy,Rodrigo:1996ha,Bilenky:1998nk,Rodrigo:1999qg}  
is presented.  

\section{Jet rates for heavy quarks}

A clustering according to the relative transverse momenta has a number of 
properties that minimize the effect of hadronization corrections and allow 
an exponentiation of leading (LL) and next-to-leading logarithms 
(NLL)~\cite{Catani:1991hj} stemming from soft and collinear 
emission of secondary partons. Jet rates in $k_\perp$ algorithms can be 
expressed, up to NLL accuracy, via integrated splitting 
functions and Sudakov form factors~\cite{Catani:1991hj}. 
For a better description of the jet properties, however, the matching 
with fixed order calculations is mandatory.
Such a matching procedure was first defined for event shapes in~\cite{Catani:1992ua}.
Later applications include the matching of fixed-order and resummed
expressions for the four-jet rate in $e^+e^-$ annihilation into massless 
quarks~\cite{Dixon:1997th,Nagy:1998bb}. A similar scheme for the matching
of tree-level matrix elements with resummed expressions in the framework
of Monte Carlo event generators for $e^+e^-$ processes was suggested 
in~\cite{Catani:2001cc} and extended to general collision types 
in~\cite{Krauss:2002up}. 

We shall recall here the results obtained in~\cite{Krauss:2003cr} for 
heavy quark production in $e^+e^-$ annihilation.
In the quasi-collinear limit~\cite{Catani:2000ef,Catani:2002hc}, the squared amplitude at 
tree-level fulfils a factorization formula, where the splitting functions $P_{ab}$ for the 
branching processes $a\to b+c$, with at least one of the partons being a heavy quark,
are given  by 
\begin{eqnarray}
P_{QQ}(z,q) \!\!\!\! &=& \!\!\!\! C_F \Bigg[ \frac{1+z^2}{1-z}- 
                   \frac{2z(1-z)m^2}{q^2+(1-z)^2 m^2}\Bigg]~, \nnb \\
P_{gQ}(z,q) \!\!\!\! &=& \!\!\!\! T_R \left[ 1 - 2z(1-z) +
                      \frac{2z(1-z)m^2}{q^2+m^2} \right]~,  
\label{eq:PQQ}
\end{eqnarray}
where $z$ is the usual energy fraction of the branching, and 
$q^2$ is the space-like transverse momentum. 
As expected, these splitting functions match the massless splitting 
functions in the limit $m\to 0$ for $q^2$ fixed. 
The splitting function 
\begin{eqnarray}
P_{gg}(z) &=& C_A \left[ \frac{z}{1-z} + \frac{1-z}{z}
                    + z(1-z) \right]~.
\end{eqnarray}
obviously does not get mass corrections at the lowest order. 

Branching probabilities are defined through~\cite{Krauss:2003cr}
\bea\label{full}
& & \!\!\!\!\!\!\!\!\!\!\!\!
\Gamma_Q(Q,q,m) = \int\limits_{q/Q}^{1-q/Q} dz 
\frac{q^2}{q^2+(1-z)^2m^2} \,P_{QQ}(z,q) \nnb \\
& & \quad = \Gamma_Q(Q,q,m=0) + 
C_F\Bigg[\frac12 - \frac{q}{m} \arctan \left(\frac{m}{q}\right) \nnb \\ 
& & \quad -\frac{2m^2-q^2}{2m^2} \log \left(\frac{m^2+q^2}{q^2}\right)
\Bigg]~,  \nnb \\           
& & \!\!\!\!\!\!\!\!\!\!\!\!
\Gamma_f(Q,q,m) = \int\limits_{q/Q}^{1-q/Q} dz 
\frac{q^2}{q^2+m^2} \,P_{gQ}(z,q) \nnb \\
& & \quad = T_R\,\frac{q^2}{q^2+m^2}\, \left[1 - \frac13\frac{q^2}{q^2+m^2}\right]~, \nnb \\  
& & \!\!\!\!\!\!\!\!\!\!\!\!
\Gamma_g(Q,q) = \int\limits_{q/Q}^{1-q/Q} dz \, P_{gg}(z)
                   = 2C_A\left(\log\frac{Q}{q}-\frac{11}{12}\right)~, 
\eea
with 
\bea\label{Gamdef}
\Gamma_Q(Q,q,m=0) &=& 2C_F\left(\log\frac{Q}{q}-\frac34\right)~,  
\eea
and the Sudakov form factors, which yield the probability for a parton 
experiencing no emission of a secondary parton between transverse 
momentum scales $Q$ down to $Q_0$, read
\bea\label{Suddef}
& & \!\!\!\!\!\!\!\!\!\!\!\!
\Delta_Q(Q,Q_0) = \exp\Bigg[-\int\limits_{Q_0}^Q 
                    \frac{dq}{q}\frac{\alpha_s(q)}{\pi}\Gamma_Q(Q,q)\Bigg] 
\,,\nnb\\
& & \!\!\!\!\!\!\!\!\!\!\!\!
\Delta_g(Q,Q_0) = \nnb \\ & & \qquad \exp\Bigg[-\int\limits_{Q_0}^Q 
                              \frac{dq}{q}\frac{\alpha_s(q)}{\pi} 
                              \left(\Gamma_g(Q,q)+
                              \Gamma_f(Q,q)\right)\Bigg]         
\,,\nnb\\
& & \!\!\!\!\!\!\!\!\!\!\!\!
\Delta_f(Q,Q_0) = \left[\Delta_Q(Q,Q_0)\right]^2/\Delta_g(Q,Q_0)\,,
\eea
where $\Gamma_f(Q,q)$ accounts for the number $n_f^{(l,h)}$ of active light or heavy 
quarks. Jet rates in the $k_\perp$ schemes can be expressed by the former branching  
probabilities and Sudakov form factors. For the two-, three- and four-jet rates
\bea \label{jetrates}
\Rate_2 &=& \left[\Delta_Q(Q,Q_0)\right]^2 
\,, \nnb \\
\Rate_3 &=& 2\left[\Delta_Q(Q,Q_0)\right]^2\, \nnb \\ &\times&
            \int\limits_{Q_0}^Q\frac{dq}{q}\frac{\alpha_s(q)}{\pi}
                               \Gamma_Q(Q,q)\Delta_g(q,Q_0)
\,, \nnb 
\nnb\\
\Rate_4 &=& 2\left[\Delta_Q(Q,Q_0)\right]^2\, \nnb \\ &\times&
            \Bigg\{\Bigg[\int\limits_{Q_0}^Q\frac{dq}{q}\frac{\alpha_s(q)}{\pi}
                         \Gamma_Q(Q,q)\Delta_g(q,Q_0)\Bigg]^2
\nnb\\
        &+&  
            \int\limits_{Q_0}^Q \frac{dq}{q} 
                  \Bigg[ \frac{\alpha_s(q)}{\pi}\Gamma_Q(Q,q)\Delta_g(q,Q_0) \nnb \\ & & \times
                         \int\limits_{Q_0}^q \frac{dq'}{q'} 
                         \frac{\alpha_s(q')}{\pi}\Gamma_g(q,q')
                              \Delta_g(q',Q_0)\Bigg]  \nnb\\
        &+ &  
            \int\limits_{Q_0}^Q \frac{dq}{q} 
                  \Bigg[ \frac{\alpha_s(q)}{\pi}\Gamma_Q(Q,q)\Delta_g(q,Q_0) \nnb \\ & & \times
                         \int\limits_{Q_0}^q \frac{dq'}{q'} 
                         \frac{\alpha_s(q')}{\pi}\Gamma_f(q,q')
                              \Delta_f(q',Q_0)\Bigg] \Bigg\}\,,
\eea
where $Q$ is the c.m. energy of the colliding $e^+e^-$, and  $Q_0^2 = y_{\rm cut} Q^2$ 
plays the role of the jet resolution scale. Single-flavour jet rates 
in Eq.~(\ref{jetrates}) are defined from the flavour of the primary vertex, 
i.e. events with gluon splitting into heavy quarks where the gluon has been emitted off 
primary light quarks are not included in the heavy jet rates but 
would be considered in the jet rates for light quarks. 

In order to catch which kind of logarithmic corrections are resummed with 
these expressions it is illustrative to study the above formulae in the 
kinematical regime such that $Q\gg m\gg Q_0$.
Expanding in powers of $\alpha_s$, jet rates can formally be expressed as
\bea
\Rate_n = \delta_{n2} +
          \sum\limits_{k=n-2}^\infty 
          \left( \frac{\alpha_s(Q)}{\pi} \right)^k\, 
          \sum\limits_{l=0}^{2k} c^{(n)}_{kl},
\eea
where the coefficients $c^{(n)}_{kl}$ are polynomials of order $l$ in 
$L_y = \log(1/y_{\rm cut})$ and $L_m = \log(m^2/Q_0^2)$. 
The coefficients for the first order in $\alpha_s$ are given by
\bea\label{expand1}
c^{(2)}_{12} = -c^{(3)}_{12} &=& -\frac12 C_F (L_y^2-L_m^2)\,,\nnb\\
c^{(2)}_{11} = -c^{(3)}_{11} &=& \frac32 C_F L_y + \frac12 C_F L_m\,.
\label{coef1}
\eea
For second order $\alpha_s$, with $n$ active flavours at the high scale,
the LL and NLL coefficients read
\bea\label{expand2}
c^{(2)}_{24} &=& \frac18 C_F^2 \left(\vp L_y^2 - L_m^2\right)^2\,,\nnb\\
c^{(3)}_{24} &=& -\frac14 C_F^2 \left(\vp L_y^2 - L_m^2\right)^2 \nnb \\ 
             &-& \frac{1}{48} C_FC_A\left(\vp L_y^4-L_m^4\right)\,,\nnb\\
c^{(4)}_{24} &=& \frac18 C_F^2 \left(\vp L_y^2 - L_m^2\right)^2  \nnb \\
             &+& \frac{1}{48} C_FC_A\left(\vp L_y^4-L_m^4\right)\,,\nnb\\\nnb\\
c^{(2)}_{23} &=& -C_F^2 \left(\vp L_y^2-L_m^2\right)
                 \left(\frac34 L_y - \frac14 L_m\right) \nnb \\ 
                       &-&\frac13\beta_n C_F 
                 \left(L_y^3-\frac32 L_yL_m^2+\frac12L_m^3\right)\nnb\\
             &-& \frac13 \left(\beta_n-\beta_{n-1}\right) C_F L_m^3 \,,\nnb\\
c^{(3)}_{23} &=& \frac12 C_F^2  \left(\vp L_y^2-L_m^2\right)
                    \left(3 L_y - L_m\right) \nnb \\  
             &+& \frac12\beta_n C_F L_y \left(\vp L_y^2 - L_m^2\right) \nnb \\   
             &+& \frac{1}{24} C_F C_A\left(\vp 3 L_y^3 - L_m^3\right) \nnb\\
             &+& \frac16 \left(\beta_n-\beta_{n-1}\right) C_F L_m
                           \left(L_y^2 - L_y L_m + 2 L_m^2\right)\,,\nnb\\
c^{(4)}_{23} &=& -C_F^2 \left(\vp L_y^2-L_m^2\right)
                         \left(\frac34 L_y - \frac14 L_m\right) \nnb \\  
             &-& \frac16\beta_n C_F \left(\vp L_y^3-L_m^3\right) \nnb \\
             &-& \frac18 C_F C_A\left(\vp L_y^3-\frac13 L_m^3\right) \nnb\\
             &-& \frac16 \left(\beta_n-\beta_{n-1}\right) C_F L_yL_m
                           \left(L_y - L_m\right)\,. 
\label{coef2}
\eea                           
Terms $\sim (\beta_{n} - \beta_{n-1})$ in the NLL coefficients, where  
the $\beta$-function $\beta_n$ for $n$ active quarks is given by
\bea
\beta_n = \frac{11\, C_A - 2\, n}{12}~,
\eea
are due to the combined effect of the gluon splitting into massive quarks and of the 
running of $\alpha_s$ below the threshold of the heavy quarks, with a corresponding change 
in the number of active flavours. With our definition of jet rates with primary 
quarks the jet rates add up to one at NLL accuracy. This statement is 
obviously realized in the result above order by order in $\alpha_s$.

The corresponding massless result~\cite{Catani:1991hj} is obtained from 
Eqs.~(\ref{coef1}) and~(\ref{coef2}) by setting $L_m \to 0$. Notice that 
Eqs.~(\ref{coef1}) and~(\ref{coef2}) are valid only for $m \gg Q_0$ and therefore 
$m\to 0$ does not reproduce the correct limit, which has to be smooth as given by 
Eq.(\ref{jetrates}). Let us also mention that for $Q \gtrsim  m$ there is a strong 
cancellation of leading logarithms and therefore subleading effects become more 
pronounced.

An approximate way of including mass effects in massless calculations, that
is sometimes used, is the ``dead cone''~\cite{Dokshitzer:fd} approximation. 
The dead cone relies on the observation that, at leading logarithmic order, 
there is no radiation of soft and collinear gluons off heavy quarks. This effect 
can be easily understood from the splitting function $P_{QQ}$ in Eq.~(\ref{eq:PQQ}). 
For $q \ll (1-z) \  m$ this splitting function is not any more enhanced at 
$z\to 1$. This can be expressed via the modified integrated splitting function
\bea\label{deadcone}
\Gamma_Q^{\rm d.c.}(Q,q,m) &=& 
\Gamma_Q(Q,q,m=0) \nnb \\ &+& 2 C_F\log\left(\frac{q}{m}\right) \Theta(m-q)\,.
\eea
To obtain this result the massless splitting function has been used, which is 
integrated with the additional constraint $z>1-q/m$. We also compare our results 
with this approximation. 

\section{Numerical results and comparison with fixed order calculations}

\begin{figure*}
\includegraphics[width=8.5cm,height=7.5cm]{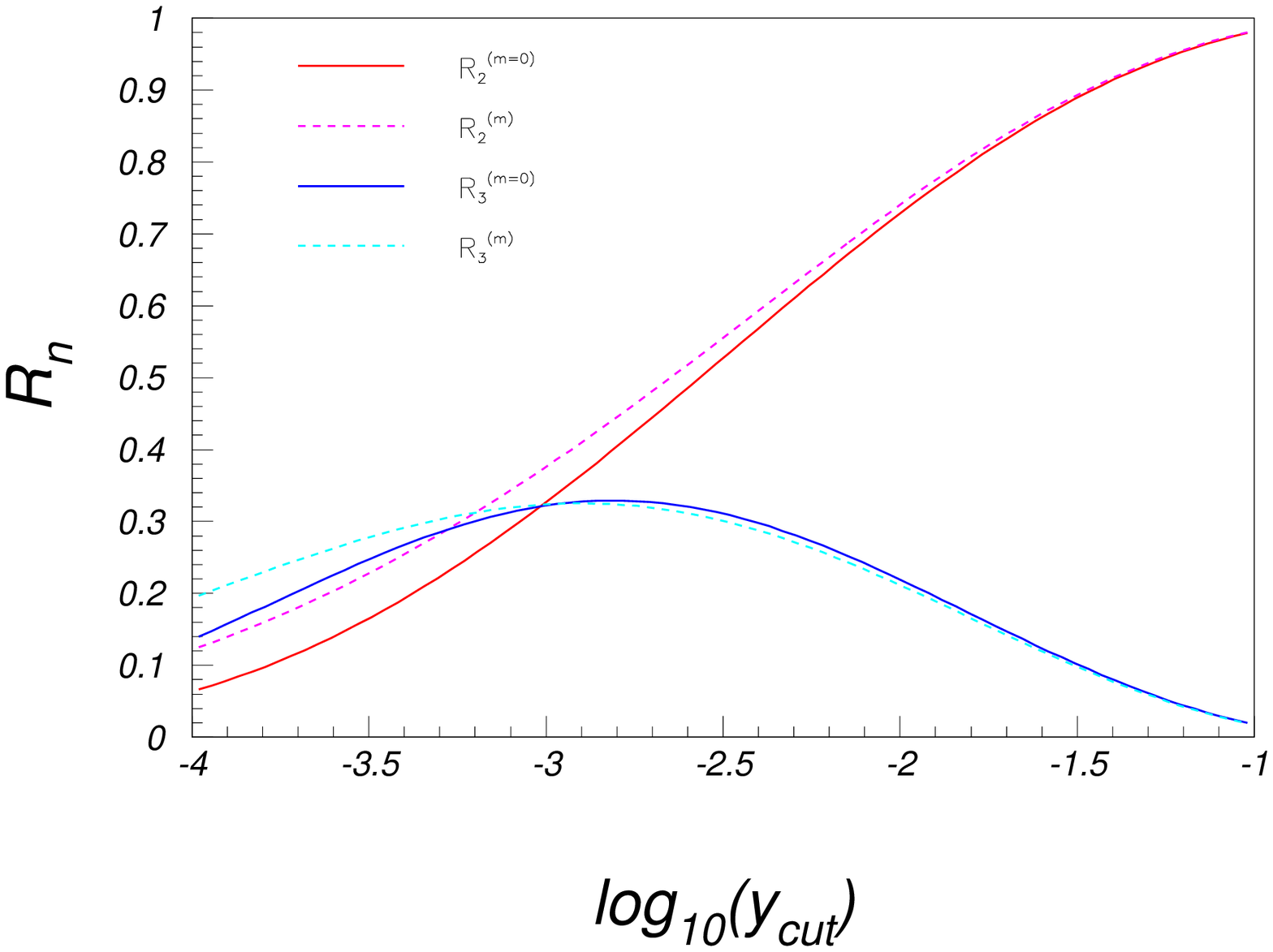} 
\includegraphics[width=8.5cm,height=7.5cm]{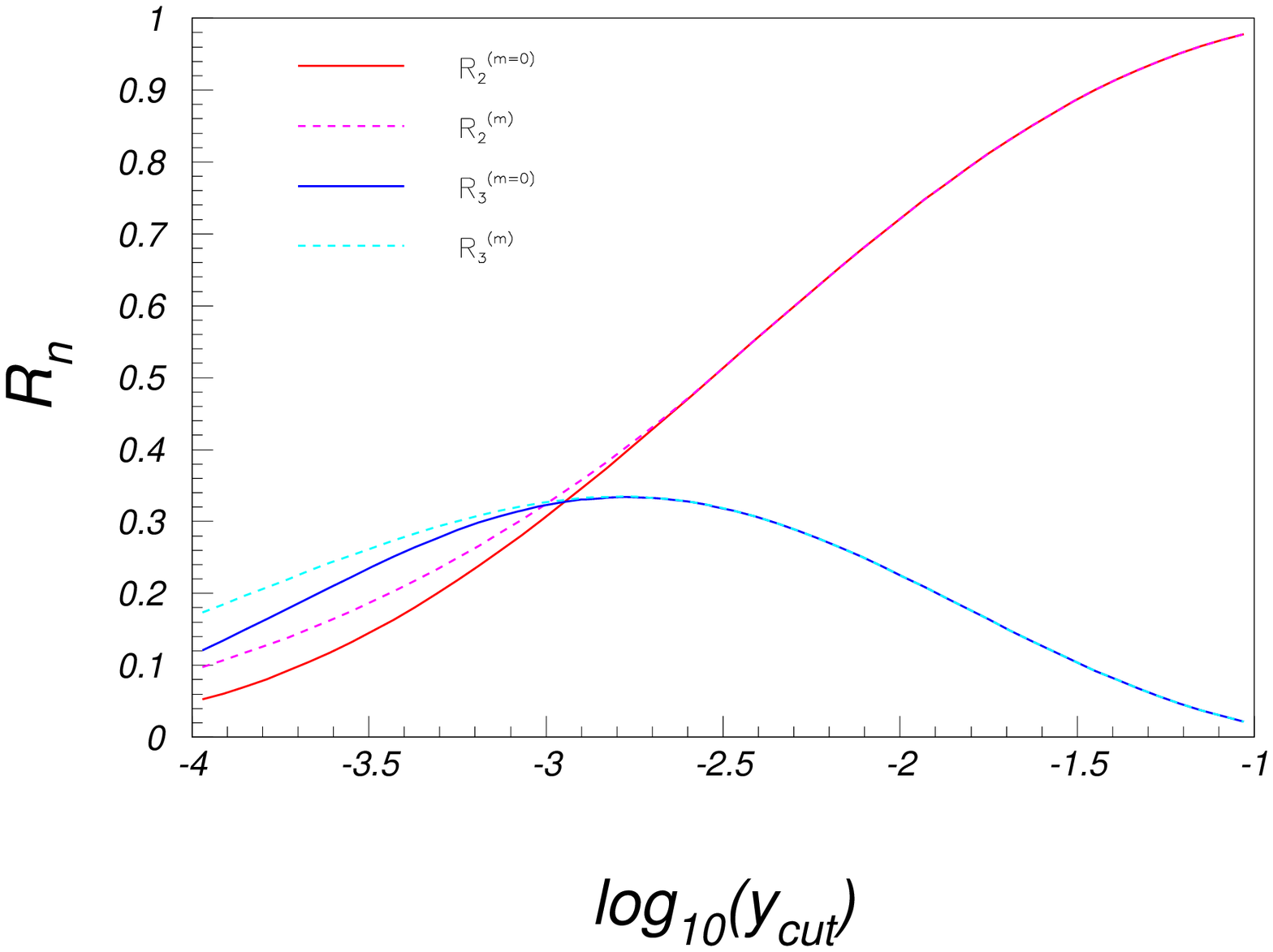}
\caption{\label{BB91} 
         Effect of a $b$-mass of $5$ GeV on the single-flavour
         two- and three-jet rate at LEP1 energies as a function of the jet 
         resolution parameter in the $k_\perp$ scheme.
         In the left plot this effect is
         treated through the full inclusion of masses into the 
         splitting function, see Eq.~(\ref{full}), whereas in the plot 
         on the right hand side this effect is modeled through the 
         dead cone, see Eq.~(\ref{deadcone}).}
\end{figure*}

\begin{figure*}
\includegraphics[width=8.5cm,height=7.5cm]{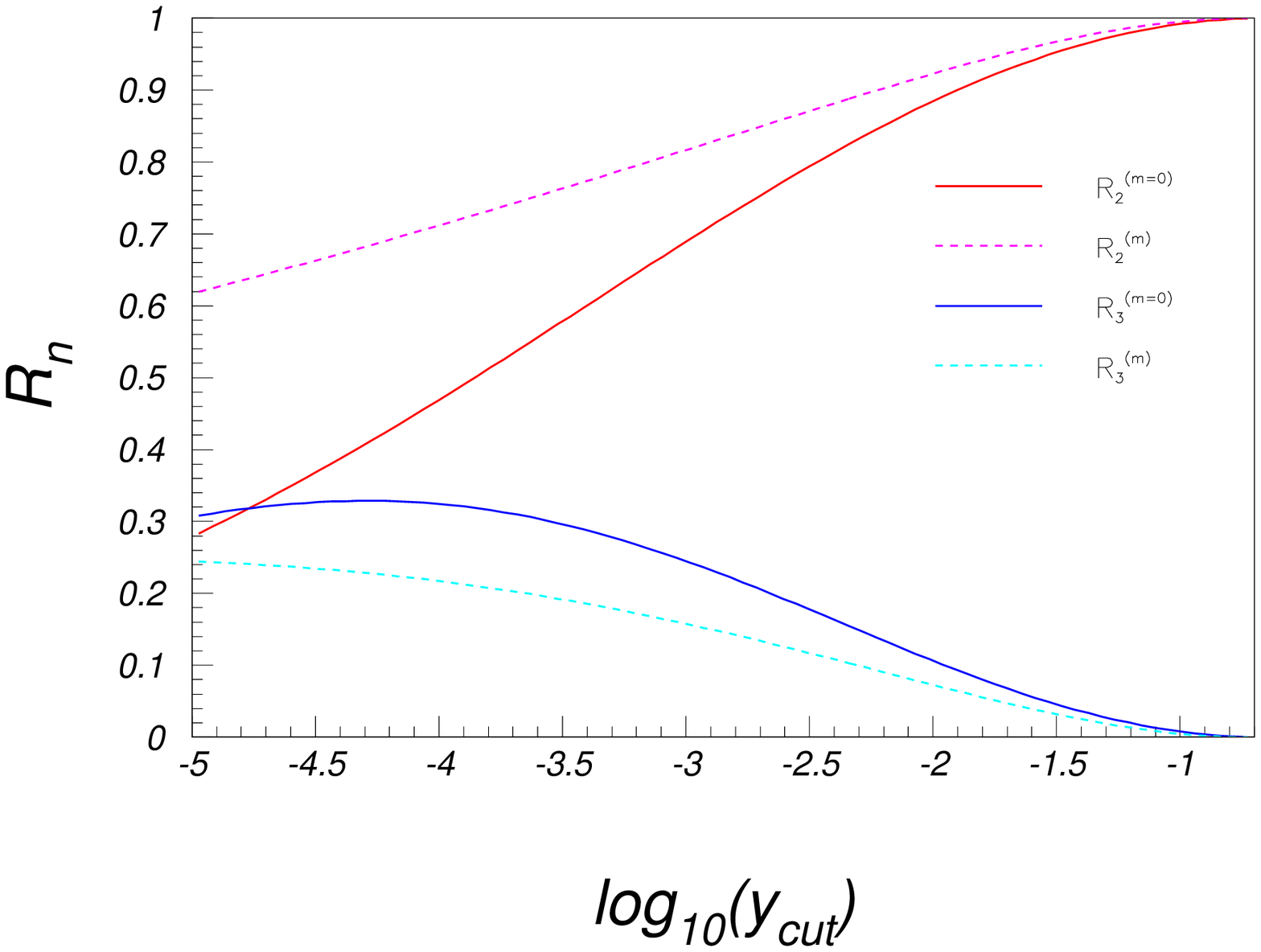} 
\includegraphics[width=8.5cm,height=7.5cm]{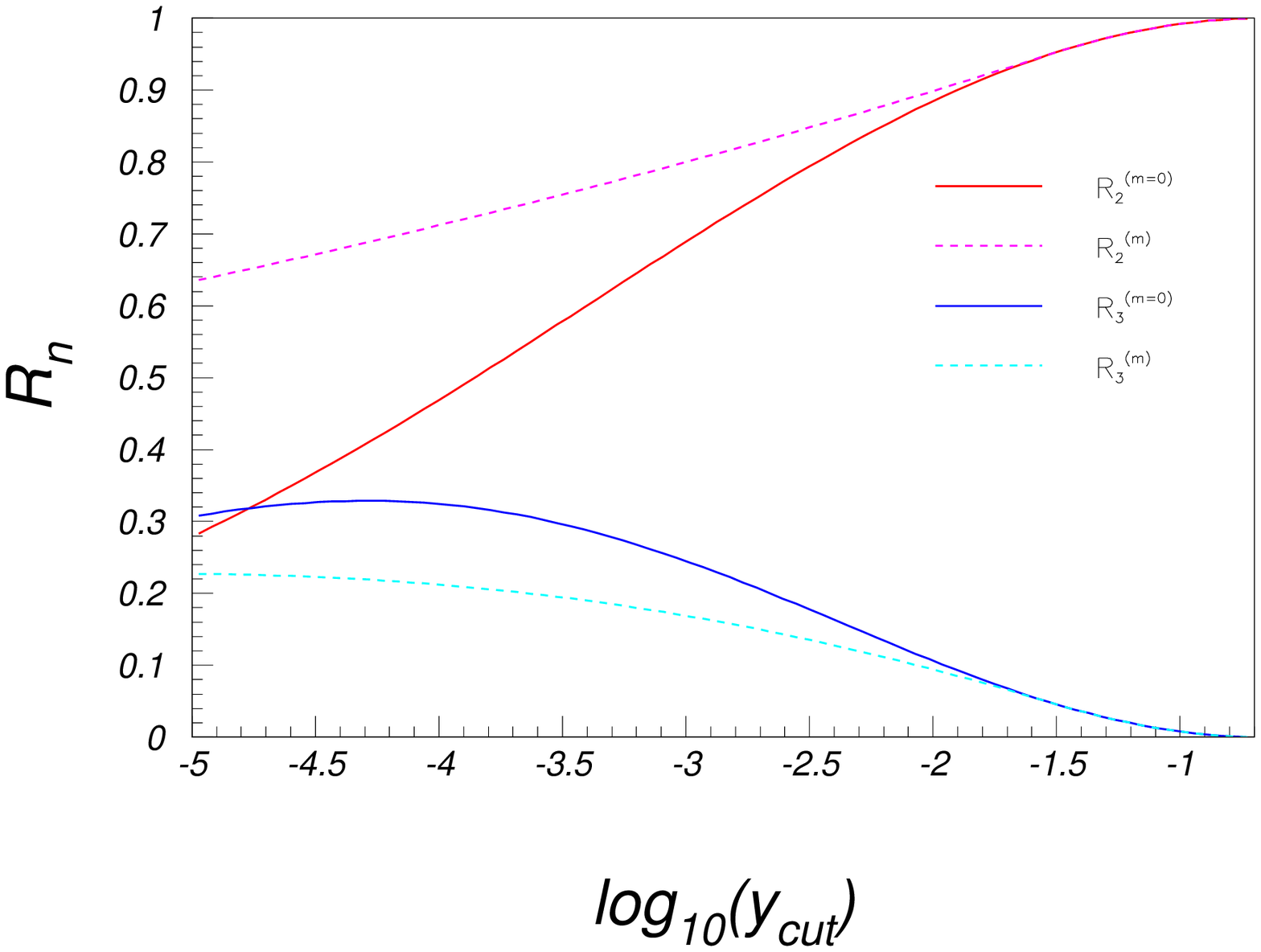}
\caption{\label{T1} 
         The effect of a $t$-mass of $175$ GeV on the single-flavour two- and 
         three-jet rate as a function of the jet resolution parameter, at a 
         potential Linear Collider operating at c.m. energies of 1~TeV.
         Again, in the left plot this effect is treated through the full 
         inclusion of masses into the splitting function, see 
         Eq.~(\ref{full}), whereas in the 
         plot on the right hand side this effect is modeled through the 
         dead cone, see Eq.~(\ref{deadcone}).}
\end{figure*}

\begin{figure*}
\includegraphics[width=9cm,height=7.5cm]{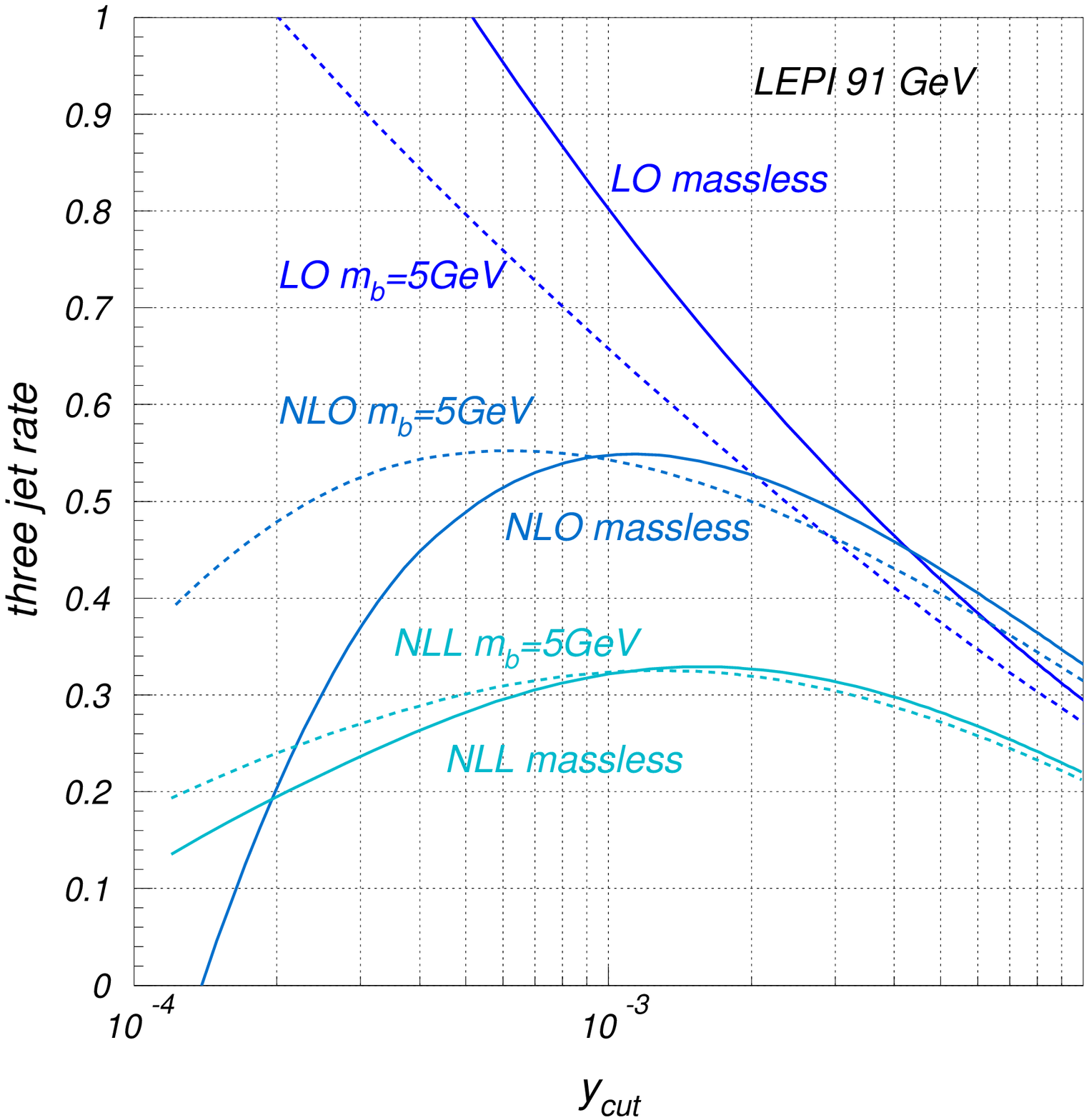} 
\includegraphics[width=9cm,height=7.5cm]{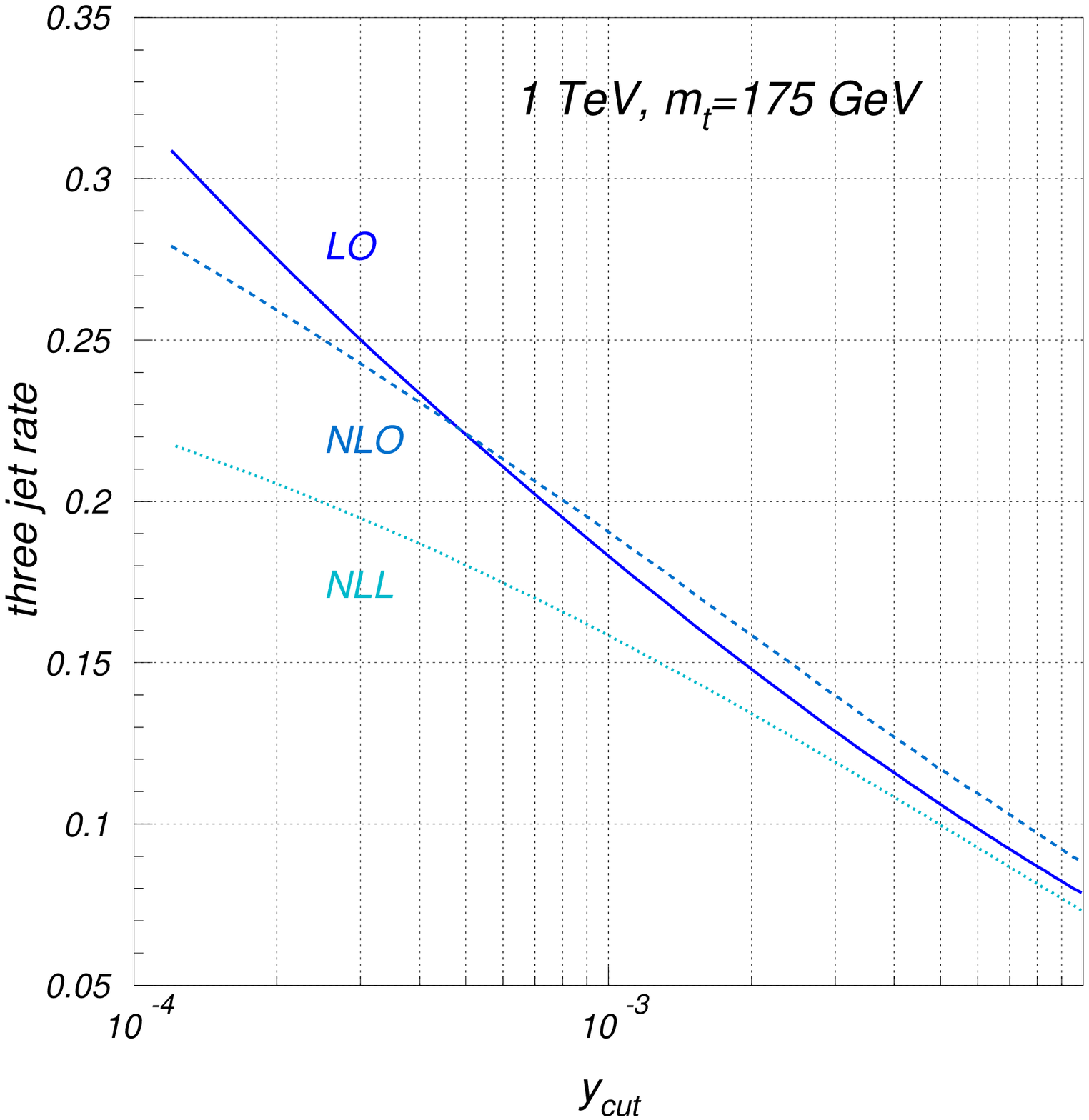} 
\caption{\label{B91} 
         Comparison of LO, NLO and NLL predictions for the single-flavour
         three-jet rate as a function of the jet 
         resolution parameter in the Cambridge algorithm 
         for bottom quark production at LEP1 energies (left plot),
         and top quark production at a potential Linear Collider operating 
         at c.m. energies of 1~TeV (right plot).}
\end{figure*}

The impact of mass effects can be highlighted by two examples, namely by
the effect of the bottom quark mass in $e^+e^-$ annihilation at the $Z$-pole, 
and by the effect of the top quark mass at a potential 
Linear Collider operating in the TeV region.

With $m_b = 5$ GeV, $M_Z = 91.2$ GeV, and $\alpha_s(M_Z) = 0.118$,
the effect of the $b$-mass at the $Z$-pole on the two- and three-jet rates 
is depicted in Fig.~\ref{BB91} (left). The result obtained in the dead 
cone approximation is shown in Fig.~\ref{BB91} (right).
Clearly, by using the full massive splitting function, 
the onset of mass effects in the jet rates is not abrupt as in the dead cone case 
and becomes visible much earlier. Already at the rather modest value of the jet 
resolution parameters of $y_{\rm cut} = 0.004$, the two-jet rate, including mass 
effects, is enhanced by roughly $4 \%$ with respect to the massless case, whereas 
the three-jet rate is decreased by roughly $3.5 \%$. For even smaller jet resolution 
parameters, the two-jet rate experiences an increasing enhancement, whereas the 
massive three-jet rate starts being larger than the massless one at values of the 
jet resolution parameters of the order of $y_{\rm cut} \approx 0.001$. The curves 
have been obtained by numerical integration of Eq.~(\ref{jetrates}). Furthermore, in 
order to obtain physical result the branching probabilities have been set to one 
whenever they exceed one or to zero whenever they become negative. 

While in the case of bottom quarks at LEP1 energies the overall effect of the 
quark mass is at the few-per-cent level, this effect becomes tremendous 
for top quarks at the Linear Collider (Fig.~\ref{T1}). 

In Fig.~\ref{B91}, leading order (LO) and next-to-leading order (NLO) predictions 
for three-jet rates are compared with the NLL result showed in the previous plots. 
Fixed order predictions for $b$-quark production clearly fail at very low values 
of $y_\mathrm{cut}$, by giving unphysical values for the jet rate, while the NLL predictions
keep physical and reveal the correct shape. The latter is an indication 
of the necessity for performing such kind of resummations. 
Fixed order predictions work well for top production at the Linear Collider,
a consequence of the strong cancellation of leading logarithmic corrections,
and are fully compatible with our NLL result.

\section{Conclusions}

Sudakov form factors involving heavy quarks have been 
employed to estimate the size of mass effects in jet rates in 
$e^+e^-$ annihilation into hadrons. These effects are sizeable and therefore 
observable in the experimentally relevant region. 
A preliminary comparison with fixed order results have been presented,
and showed good agreement.  
Matching between fixed-order calculations and resummed results
is in progress~\cite{inprogress}.

\section*{Acknowledgements}

It is a pleasure to thank the organizers of this meeting 
for the stimulating atmosphere created during the workshop,
and M. Mangano for very useful comments.
G.R. acknowledges partial support from 
Generalitat Valenciana under grant CTIDIB/ 2002/24 and 
MCyT under grants FPA-2001-3031 and BFM 2002-00568.

\end{document}